\documentclass[fp,twocolumn]{jpsj3}

\usepackage{graphicx}
\usepackage{bm}
\usepackage{txfonts}
\usepackage{color}
\usepackage{multirow}
\newcommand{\Rou}{Cu$_2$(OH)$_3$NO$_3$}
\newcommand{\rev}{black}

\title{NMR Determination of the Low-Field Magnetic Structure of the Cu-Based Mineral Rouaite, \Rou}
\author{Issei Niwata$^1$, R. Kumar$^1$, Aswathi Mannathanath Chakkingal$^2$, Anton A.\ Kulbakov$^2$, Maxim Avdeev$^{4,5}$, Dmytro S. Inosov$^{2,3}$, Darren C. Peets$^2$, Yoshihiko Ihara$^1$\thanks{yihara@phys.sci.hokudai.ac.jp}}
\inst{$^1$Department of Physics, Faculty of Science, Hokkaido University, Sapporo 060-0810, Japan \\
$^2$Institut f\"ur Festk\"orper- unt Materialphysik, Technische Universit\"at Dresden, 01062 Dresden, Germany\\
$^3$W\"urzburg-Dresden Cluster of Excellence on Complexity and Topology in Quantum Matter\,---\,\textit{ct.qmat}, Technische Universit\"at Dresden, 01062 Dresden, Germany\\
$^4$Australian Nuclear Science and Technology Organisation, Sydney, New South Wales 2234, Australia\\
$^5$School of Chemistry, The University of Sydney, Sydney, New South Wales 2006, Australia}

\abst{\noindent
Frustrated interactions in the Cu-based mineral rouaite with alternating antiferromagnetic and ferromagnetic spin chains, \Rou, introduce non-trivial magnetic ground states and exotic excitations arising from them.  
We investigated the magnetic structure of \Rou\ by $^1$H- and $^2$H-NMR measurements on single crystals. 
The internal fields in the ordered state were microscopically measured using the H nuclear moments as a local probe. 
The directions of the ordered moments were determined by comparing the experimental results to model calculations.  
The obtained magnetic structure suggests the importance of Dzyaloshinskii-Moriya interactions in stabilizing the low-field magnetic structure. 
The present result advances the theoretical understanding of the low-field magnetic states and will enable exploration of the exotic magnetic states emerging in high magnetic fields. 
}

\begin{document}
\maketitle

\section{Introduction}
In a magnetic ground state, the energy associated with the interacting magnetic moments is minimized by optimizing the magnetic structure. 
A nontrivial magnetic state appears when competing interactions frustrate magnetic moments, preventing all interactions from being simultaneously optimized. 
A typical example is antiferromagnetically coupled spins arranged on a triangle-based magnetic network, in which the three spins on the corners of the triangle cannot all be mutually anti-aligned.
Among many exotic ground states in such a system, quantum spin liquids have been theoretically suggested for the triangular lattice \cite{Anderson-1973} and have been reported in quantum magnets hosting triangular \cite{Paddison-Nphys2017} and kagome networks \cite{Shores-JACS127, Han-Nature492}.

A more complex magnetic state can be generated in a crystalline lattice with lower symmetry, in which the antiferromagnetic (AFM) and ferromagnetic (FM) interactions coexist on crystallographically independent crystallographic sites in a unit cell.  
The competing interactions prohibit long-range magnetic order even at temperatures much lower than the energy scale of the interactions and introduce exotic magnetic states assisted by the quantum fluctuations which become relevant at low temperatures. 
Quantum effects are most apparent for $S = 1/2$ quantum spins on a low-dimensional network, where the reduced number of interaction pathways lowers the stability of any long-range order. 
An AFM chain with $S=1/2$ spins hosts a Tomonaga--Luttinger liquid state \cite{Giamarchi-2004} and generates fractionalized spin excitations (spinons) \cite{Haldane-PRL66}, while the elemental excitations of the FM chain are magnons. 
By combining AFM and FM chains in a single unit cell, coexisting and interacting spinons and magnons were reported in the Br variant of botallackite, Cu$_2$(OH)$_3$Br \cite{Zhang-PRL125}.  To date, no other systems have been reported to host spinons and magnons in the same energy range.  
To reveal the magnetic states caused by the interaction of bosonic and fermionic particles, 
magnetic materials having coupled AFM and FM chains have been investigated \cite{Zheng-PBCM404,Zhao-JPCM31, Fujita-CA7, Zheng-PRB71, Fujita-ACS15,Wulferding-PRB111,Kulbakov-PRB106}.

\begin{figure}
\includegraphics[width=\columnwidth]{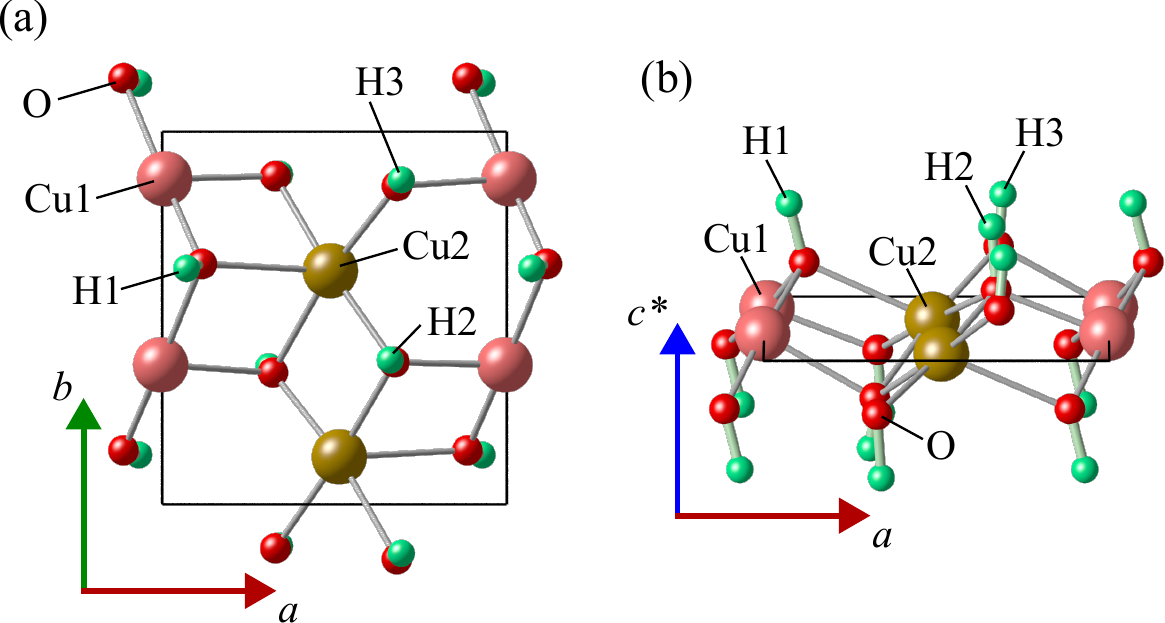}
\caption{
  Crystal structure of \Rou: (a) projected on the $\bm{ab}$ plane and (b) viewed from a direction close to the $\bm{b}$ axis.  
  The thin black lines represent the unit cell. 
  Cu1 (Cu2) sites form FM (AFM) chains along the $\bm{b}$ direction through Cu--O--Cu superexchange paths. 
  The chain direction coincides with the $2_1$ screw axis in the $P2_1$ space group. 
  H1, H2, and H3 sites are connected to O sites to form OH$^-$ groups, and participate in an interlayer hydrogen-bonding network essential for stabilizing the structure but not expected to allow significant exchange interactions. 
  These hydrogen bonds are to NO$_3^{-}$ ions (not shown) stacked above and below the 2-D magnetic network, leading to a long and tortuous magnetic exchange pathway along the $\bm{c}$ direction.  
}
\label{Structure}
\end{figure}

The Cu-based quantum magnet \Rou\, (mineral name: rouaite) has a crystal structure similar to that of botallackite, namely a two-dimensional (2-D) magnetic network composed of alternating AFM and FM chains. \cite{Effenberger-ZKCM165} 
The crystal structure of the magnetic plane is illustrated in Fig.~\ref{Structure}. 
The 2-D planes are separated by nonmagnetic NO$_3^-$ ions in rouaite in place of Br$^-$ in botallackite.  
The magnetism in rouaite stems from the $S=1/2$ magnetic moments of Cu$^{2+}$ ions with $3d^9$ electronic configuration.
The dominant Cu1-Cu1 and Cu2-Cu2 superexchange interactions through oxygen ligands lead to quasi one-dimensional (1-D) chains along the $\bm{b}$ direction. 
Theoretical work suggests FM exchange interactions along the Cu1 chains and AFM interactions along the Cu2 chains \cite{Ruiz-JPCB110,Pillet-PRB73, Kikuchi-JPCS969}. 
The AFM and FM chains are connected by weaker inter-chain interactions, which are expected to be frustrated, and the inclusion of even weaker inter-plane interactions induces three-dimensional (3-D) magnetic order in the material. 
Magnetization and heat capacity measurements show anomalies associated with successive phase transitions at $T_\text{N1}$ and $T_\text{N2}$, suggesting complex magnetic states caused by a shifting equilibrium among competing magnetic interactions \cite{Kikuchi-JPCS969,Yuan-PRB106, Chakkingal-PRB110}. 
The application of a magnetic field introduces further nontrivial magnetic states by modifying the balance between AFM and FM interactions. 
As a result, multiple magnetic phases are found in the temperature-field phase diagram constructed from measurements of several physical quantities \cite{Chakkingal-PRB110}. 
To understand the intriguing magnetic excitations in this combination of FM and AFM chains in \Rou, the magnetic structure of the ground state should be studied by microscopic measurements.

The magnetic structure in the zero field has been investigated by neutron diffraction measurements \cite{Yuan-PRB106,Chakkingal-PRB110}. 
The magnetic translation vector was determined to be $\bm{Q}=(0.5,0,0)$ from the magnetic Bragg peaks observed in highly $^2$H-enriched samples, and the magnetic order was refined \cite{Chakkingal-PRB110}.  
However, the weak scattering from the $S=1/2$ quantum spins and the high incoherent scattering background from residual $^1$H led to considerable uncertainty in the spins' canting angles.  
To further refine the magnetic structure, we performed $^1$H- and $^2$H-NMR spectroscopic measurements at low magnetic fields on a single crystal of partially-deuterated \Rou. 
The directions of the magnetic moments in the long-range-ordered state are precisely determined by our NMR measurements, which use H nuclear moments as the magnetic local probe.  
The internal fields generated in the magnetically ordered state were quantitatively estimated from the NMR spectra and were compared to simulations based on the magnetic structure model from diffraction. 
The magnetic interactions present in \Rou, which are crucial to orienting the ordered moments, are discussed on the basis of the obtained magnetic structure.  

\begin{figure*}[t]
\begin{center}
\includegraphics[width=0.95\textwidth]{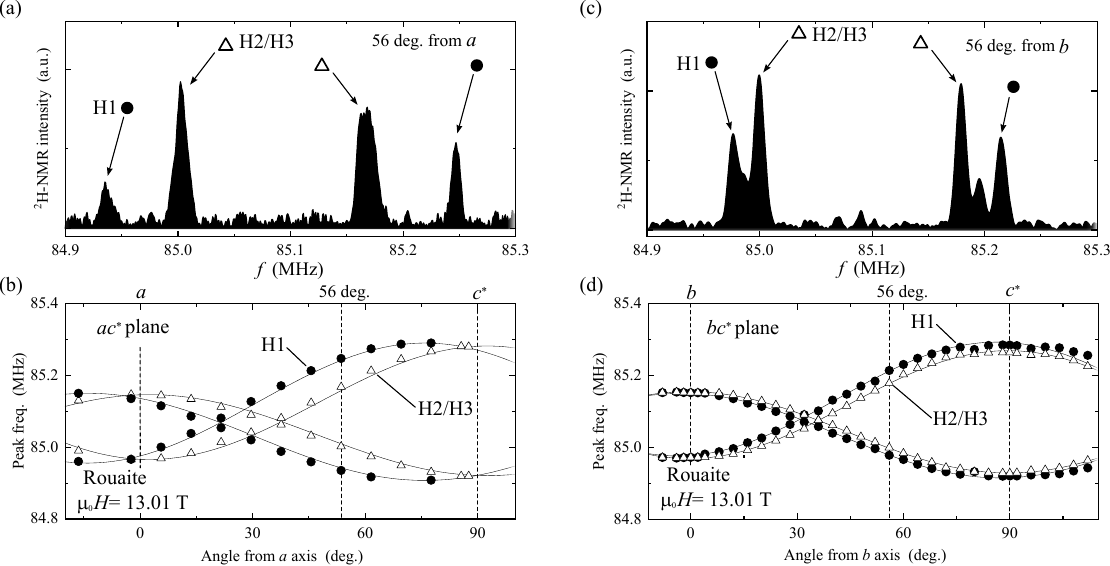}
\caption{
  (a), (c) $^2$H-NMR spectra at room temperature measured in fields applied at 56$^\circ$ from the (a) $\bm{a}$ and (c) $\bm{b}$ axes. 
  The NMR peaks assigned to the H1 and H2/H3 sites are labeled by filled circles and open triangles, respectively. 
  (b), (d) Angle dependence of peak positions measured upon rotation within the (b) $\bm{ac}^{\ast}$ and (d) $\bm{bc}^{\ast}$ planes. 
  The circle and triangle symbols correspond to the NMR peaks assigned in (a) and (c).  
  The field angle was measured from the $\bm{a}$ axis for (b) and from the $\bm{b}$ axis for (d). 
  The field orientations for the NMR spectra shown in (a) and (c) are indicated by vertical dashed lines, together with the $\bm{a}$, $\bm{b}$ and $\bm{c}^{\ast}$ directions.  
}
\label{RTSP}
\end{center}
\end{figure*}

\section{Experimental}
Large single crystals of \Rou\ with a typical dimension of $3 \times 5 \times 10$ mm$^3$ were obtained by hydrothermal synthesis \cite{Chakkingal-PRB110}. 
Cu(NO$_3$)$_2$$\cdot$3H$_2$O was dissolved in deuterated water (D$_2$O) and heated at 240 $^{\circ}$C for five days in a PPL-lined autoclave. 
This combination of deuterated water with protons from the copper nitrate hydrate produced crystals with approximately 60\% deuteration, allowing us to obtain both $^2$H- and $^1$H-NMR spectra on the same sample. 
$^2$H and $^1$H randomly occupy the three crystallographically independent H sites, H1, H2 and H3 (Fig.~\ref{Structure}).

$^2$H-NMR spectra were measured above 100 K in a fixed field of 13.01 T to estimate the microscopic parameters used for the analyses at low temperature. 
The field strength was calibrated by the $^{13}$C NMR frequency of tetramethylsilane and the gyromagnetic ratio of $^{13}$C, $\gamma_{\rm C} = 10.5042$ MHz/T. 
In the fixed-field measurements, the NMR frequency spectrum was obtained by the Fourier transform (FT) of the spin-echo signal to optimize the spectral resolution. 
At low temperatures, NMR spectra were obtained by the field-sweep method, for which the signal intensity at a fixed measurement frequency was recorded while sweeping the field at a constant rate.
The transient FT spectra \textcolor{\rev}{were} mapped on the field axis by using the magnetic field at the moment of signal detection. 
The orientation of the single crystal in the magnetic field is controlled by a home-built single-axis rotator.

\section{Result}
\subsection{$^\text{2\!}$H-NMR measurement in the paramagnetic state}

$^2$H-NMR measurements were performed at room temperature to assign observed NMR peaks to each crystallographic H site. 
In a unit cell, H atoms occupy three different sites: H1 at $(0.1013, 0.6321, 0.2800)$, H2 at $(0.6890, 0.3871, 0.2612)$, and H3 at $(0.7215, 0.8695, 0.2659)$ \cite{Chakkingal-PRB110}. 
The H1 sites are more closely connected to the FM Cu1 chain and the H2/H3 sites are closer to the AFM Cu2 chain. 
These hydrogens form OH$^-$ groups with a nearby host oxygen through covalent bonds.  
These covalent electrons dominate the electric field gradient (EFG) at the H sites, making the main principal axis of the EFG nearly parallel to the O--H bond. 
The orientation of the EFG principal axes affects the $^2$H-NMR spectrum through the electric quadrupolar interaction. 
For the $^2$H nuclear spin with $I=1$, the nuclear-spin energy levels in a magnetic field $\bm{H}$ are characterized by the sum of the Zeeman and the electric quadrupolar interactions as 
\begin{align*}
\mathcal{H} &= -\gamma \hbar (1+K)\mu_0 \bm{H} \cdot \bm{I}  \notag + \frac{\nu_Q}{6}\left[ \left( 3I_z^2 - \bm{I}^2 \right) + \frac{1}{2}\eta \left( I_+^2+I_-^2 \right) \right]. 
\end{align*}
Here, $\gamma$, $\hbar$, $K$, and $\mu_0$ are the gyromagnetic ratio, the reduced Planck constant, Knight shift, and vacuum permeability, respectively, while $\nu_Q$ represents the nuclear quadrupolar resonance (NQR) frequency and $\eta$ is the asymmetry parameter of the EFG.   
We assume a local coordinate system for $\bm{I}$ that is aligned with the EFG axes. 
The nuclear-spin levels depend on the orientation of the external field in this coordinate system, leading to a sinusoidal variation of the NMR frequency upon rotating a single-crystalline sample in an external field, as seen in Figs.~\ref{RTSP}(b) and \ref{RTSP}(d). 

\begin{figure}
\begin{center}
\includegraphics[width=0.9\columnwidth]{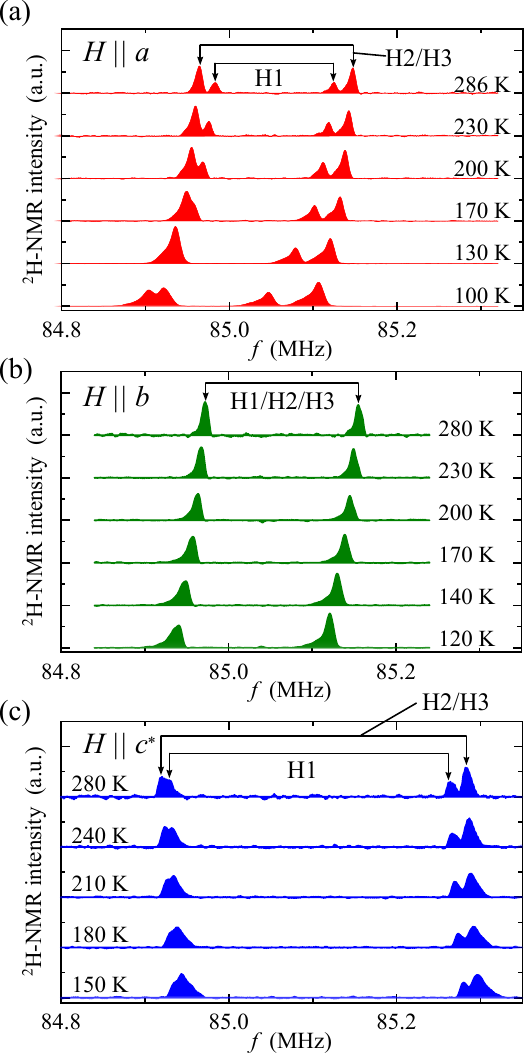}
\caption{
  $^2$H-NMR spectra at representative temperatures measured in fields along the (a) $\bm{a}$, (b) $\bm{b}$, and (c) $\bm{c}^{\ast}$ axes.  
  The peak positions shift to lower frequencies for $\bm{H}\!\parallel\bm{a}$ and $\bm{H}\!\parallel\bm{b}$ and higher frequencies for $\bm{H}\!\parallel\bm{c}^{\ast}$ due to the increase in the magnetization at low temperatures. 
  The direction of the shift is determined by the sign of the coupling constants between the nuclear and electronic magnetic moments. 
  The H1 and H2/H3 sites are more clearly resolved at low temperatures where the magnetization is larger for $\bm{H}\!\parallel\bm{a}$ and $\bm{H}\!\parallel\bm{c}^{\ast}$, while all sites are overlapped for $\bm{H}\!\parallel\bm{b}$ due to the similar coupling constants along the $\bm{b}$ axis for all H sites. 
}
\label{tdep}
\end{center}
\end{figure}

From a single $^2$H nuclear site, two NMR peaks will be observed arising from the $m=1\leftrightarrow 0$ and $m=0\leftrightarrow -1$ transitions. 
In the experimentally obtained spectrum shown in Fig.~\ref{RTSP}(a), two large and two small peaks were observed, suggesting two overlapping and one independent H sites in line with the three crystallographic H sites. 
Here, we assume that the H2 and H3 sites overlap due to their similar local environment.  
This spectrum was measured at room temperature in a fixed field of 13.01 T applied along a direction 56$^\circ$ from the $\bm{a}$ axis to ensure a good peak separation. 
The peak positions were tracked upon rotating the field orientation within the $ac^{\ast}$ plane, and the result is shown in Fig.~\ref{RTSP}(b). 
The resonant frequencies of the large and small peaks are plotted by open triangles and filled circles, as indicated in Fig.~\ref{RTSP}(a). 
The wider (narrower) peak splitting for fields along the $\bm{c}^{\ast}$ ($\bm{a}$) direction and the crossing of the branches at an intermediate angle are both consistent with the orientation of the O--H bond, and thus the main axis of EFG, being nearly parallel to the $\bm{c}^{\ast}$ direction.
A shift of 16$^\circ$ in the sinusoidal behavior of the sites reflects the relative angle between the O--H bonds for the H1 and H2/H3 sites, which was estimated to be 14$^\circ$ based on the published crystal structure refinement \cite{Chakkingal-PRB110}. 
The H positions determined by neutron diffraction are consistent with the present $^2$H-NMR results; on the other hand, the H positions generated by density functional theory \cite{Yuan-PRB106} do not agree with the present observation. 
The field-orientation dependence of the NMR peak positions was also measured in the $bc^{\ast}$ plane, as shown in Figs.~\ref{RTSP}(c) and (d). 
The relative phase shift is smaller in this plane reflecting nearly parallel O--H bonds for all three H sites within $\pm 5^\circ$. 
By fitting the sinusoidal angular dependence we obtained the NQR frequencies for the H1 and H2/H3 sites to be 380 and 350 kHz, respectively.
The small $\eta < 0.05$ \textcolor{\rev}{for all H sites} confirms the axially symmetric local environment along the O--H bonds.  
\textcolor{\rev}{As the crystallographic symmetry of the H site is lower than axial, this result suggests that the covalent electrons dominate the EFG at H sites.} 

\begin{figure}
\begin{center}
\includegraphics[width=0.9\columnwidth]{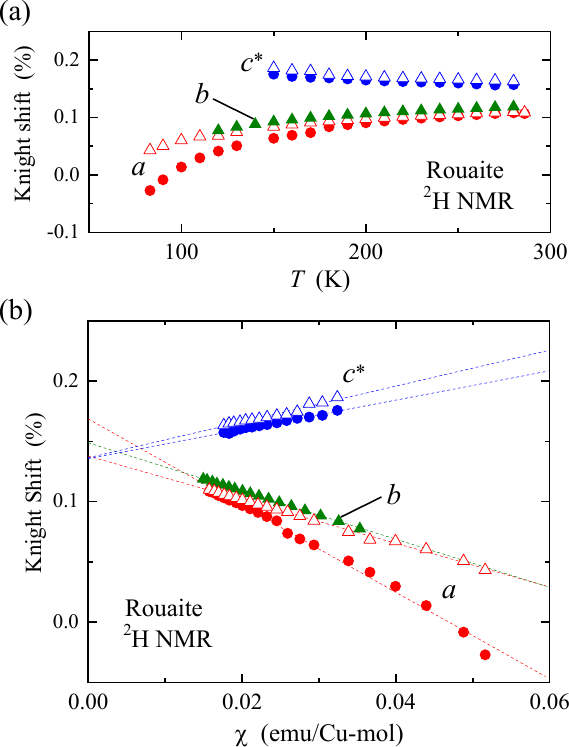}
\caption{
  (a) Temperature dependence of Knight shift for the $\bm{a}$, $\bm{b}$, and $\bm{c}^{\ast}$ directions. 
  The Knight shifts for the H1 and H2/H3 sites are represented by filled circles and open triangles for the $\bm{a}$ and $\bm{c}^{\ast}$ directions, respectively. 
  For the $\bm{b}$ direction, the Knight shift is plotted with filled triangles as the three H sites are not resolved in the NMR spectra. 
  (b) The Knight shift is plotted against the bulk magnetization at the corresponding temperatures.  
  The dashed lines are the results of linear fitting. 
  From the slopes of these lines, the hyperfine coupling constants are estimated. 
  The finite intercept at $\chi=0$ is caused by the electric quadrupolar interactions for $^2$H nuclear spins. 
}
\label{Kchi}
\end{center}
\end{figure}

\begin{table}[b]
  \caption{Coupling constants estimated in the paramagnetic state. All values are in units of mT/$\mu_{\text B}$.}
  \label{tab:Ahf}
  \centering
    \begin{tabular}{llccc}\hline
    & Site & $\bm{a}$ & $\bm{b}$ & $\bm{c}^{\ast}$ \\
                           \hline
    \multirow{2}{*}{Total} &H1     & $-200$ & $-112$ & $+68$ \\
                           & H2/H3 & $-101$ & $-112$ & $+83$ \\ 
                           \hline
    \multirow{2}{*}{Dipole} &H1     & $-75$ & $-23$ & $+98$ \\
                           & H2/H3 & $-30$ & $-40$ & $+70$ \\ 
                           \hline
    \multirow{2}{*}{Hyperfine} &H1     & $-125$ & $-89$ & $-30$ \\
                           & H2/H3 & $-71$ & $-72$ & $+13$ \\ 
                           \hline
    \end{tabular}
\end{table}

\begin{figure*}[t]
\begin{center}
\includegraphics[width=0.95\textwidth]{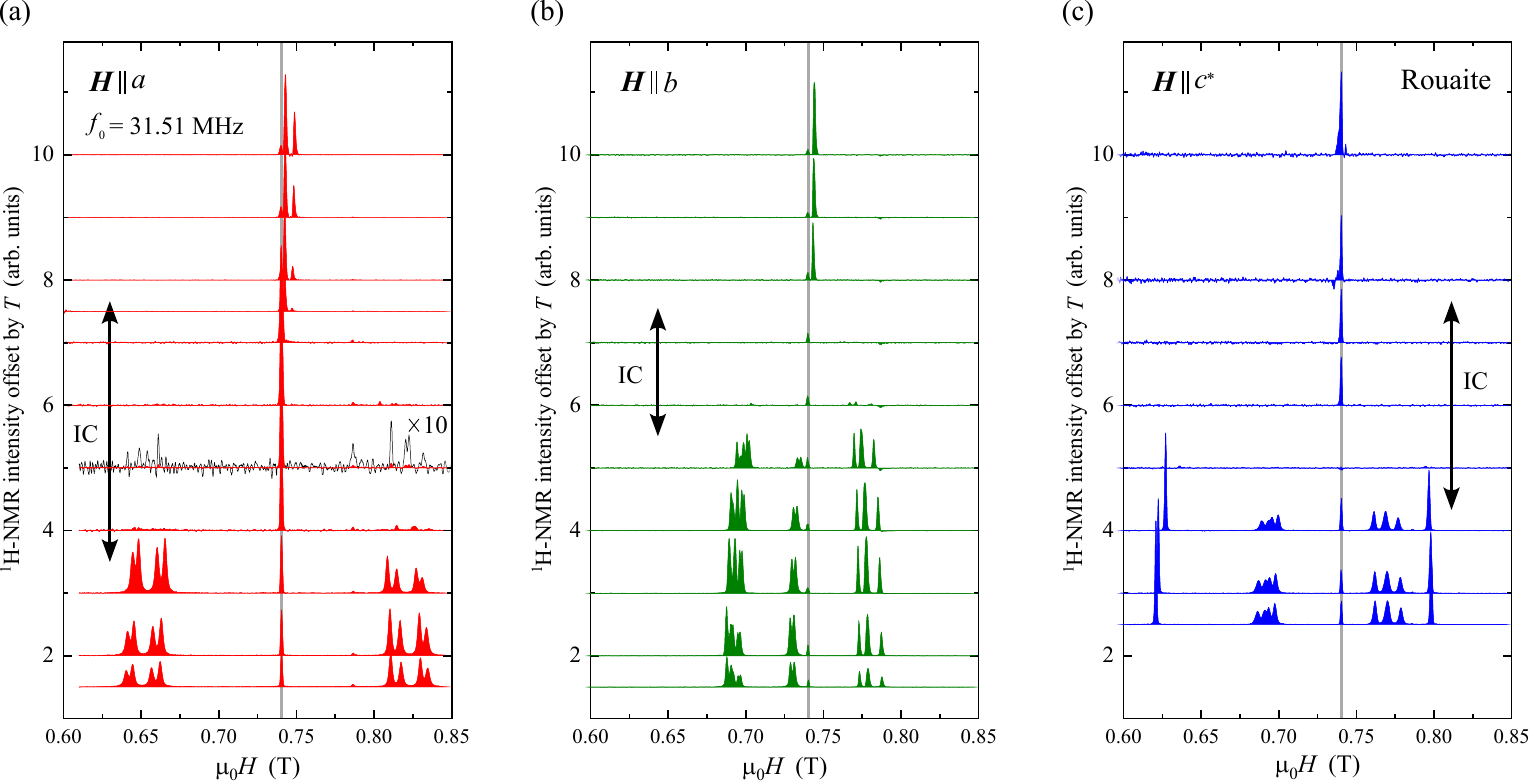}
\end{center}
\caption{
  Field-sweep $^1$H-NMR spectra for fields along the (a) $\bm{a}$, (b) $\bm{b}$, and (c) $\bm{c}^{\ast}$ directions. 
  For all the field directions, the sharp $^1$H-NMR spectra in the paramagnetic state disappear once below the main transition at $T_\text{N1}$ and are restored below $T_\text{N2}$.
  \textcolor{\rev}{A spectrum at 5 K in $H\parallel a$ is expanded by a factor of ten and shown together by a black line.}
  The transition temperatures depend on the field orientation. 
  The incommensurate (IC) phase in the intermediate temperature range between $T_\text{N1}$ and $T_\text{N2}$ is indicated by vertical arrows. 
  In the ground state below $T_\text{N2}$, many split peaks due to the internal fields were observed. 
}
\label{AFMSP}
\end{figure*}

The temperature dependence of the NMR shift was measured in the paramagnetic state above 100\,K to estimate the hyperfine coupling constants between the Cu$^{2+}$ electronic moments and the $^2$H nuclear magnetic moments.  
Figure \ref{tdep} shows the temperature variation of the $^2$H-NMR spectra for fields along the $\bm{a}$, $\bm{b}$ and $\bm{c}^\ast$ directions. 
For $\bm{H}\!\parallel\bm{a}$ and $\bm{H}\!\parallel\bm{c}^{\ast}$, the large and small peaks are assigned as H2/H3 and H1 sites following the assignment above. 
As the spectra for $\bm{H}\!\parallel\bm{b}$ are not resolved as shown in Fig.~\ref{tdep}(b), we assumed the same NMR shift for all three H sites in this field direction.  
The peak positions shift to lower frequencies on cooling for $\bm{H}\!\parallel\bm{a}$ and $\bm{H}\!\parallel\bm{b}$, but to higher frequencies for $\bm{H}\!\parallel\bm{c}^{\ast}$.  
This behavior demonstrates the anisotropic Knight shift due to the dominant dipolar contribution.

To extract $K$, the peak separation due to the electric quadrupolar interaction is eliminated by taking the average of the separated peak frequencies. 
The resulting temperature dependence of $K$ for all directions and sites is summarized in Fig.~\ref{Kchi}(a).  
The results for the H1 site are represented by filled circles and those for the H2/H3 sites are shown by open triangles. 
The results for all H sites for $\bm{H}\!\parallel\bm{b}$ are indicated by filled triangles.  
In Fig.~\ref{Kchi}(b), $K$ is plotted as a function of magnetic susceptibility $\chi$ measured at the corresponding temperatures and field directions. 
From the linear relationship $K(T) = A_{\rm total}\,\chi (T)$, the total coupling constants $A_{\rm total}$ were estimated for each H site and field direction. 
$A_{\rm total}$ is a sum of the dipole and transferred hyperfine couplings. 
The direct dipole contribution is calculated from the crystal structure by adding the dipole fields from the Cu$^{2+}$ moments within 100 \AA\, from the target H sites.
The hyperfine coupling constants are then estimated by subtracting the dipole contribution from the experimentally obtained $A_{\rm total}$, leading to the results listed in Table \ref{tab:Ahf}. 
The coupling strengths for $^1$H nuclear spins are the same as those obtained here, although the gyromagnetic ratio is different, because these are the effective fields generated by the Cu$^{2+}$ moments. 
The coupling constants are used to identify the magnetic structure in the ordered state from analyses of the $^1$H-NMR spectra. 

\subsection{$^\text{1\!}$H-NMR measurement in the ordered state}

The magnetic structure in the ordered state below $T_\text{N1}= 7.2$\,K was investigated by $^1$H-NMR spectroscopy. 
The sharp $^1$H-NMR spectra in the paramagnetic state split into several peaks in the ordered state due to the effects of the spontaneous internal fields. 
To observe all NMR peaks emerging in the ordered state, we measured the $^1$H-NMR spectra by the field-sweep method at the fixed frequency $f_0 = 31.51$ MHz. 
The resonant field for protons at $f_0$ is 0.74 T, which is small enough to remain firmly within the low-field magnetic phase. 
Figure \ref{AFMSP} summarizes the temperature dependence of the $^1$H-NMR spectra at low temperatures in fields along the $\bm{a}$, $\bm{b}$, and $\bm{c}^\ast$ directions. 
For all field orientations, a sharp and temperature-independent peak was observed at the reference field of 0.74 T as indicated by vertical gray lines. 
This peak is assigned to a background signal arising from $^1$H nuclear spins in the grease, sample stage, and insulation on the Cu wires. 
The background signal overlaps with the signal from the sample for $\bm{H}\!\parallel\bm{c}^{\ast}$ in the paramagnetic state.  
In the AFM state, however, as the internal fields significantly shift the intrinsic NMR peaks, the background signal is well separated from all sample signals and does not affect our analysis.  
The peaks from the H1 and H2/H3 sites are already well separated in the paramagnetic state for $\bm{H}\!\parallel\bm{a}$ due to the high contrast of the coupling constants for these sites. 
These H sites are less resolved for the other field orientations, as already seen in the $^2$H NMR spectra at high temperatures (Fig.~\ref{tdep}). 

\begin{figure}[t]
\begin{center}
\includegraphics[width=0.9\columnwidth]{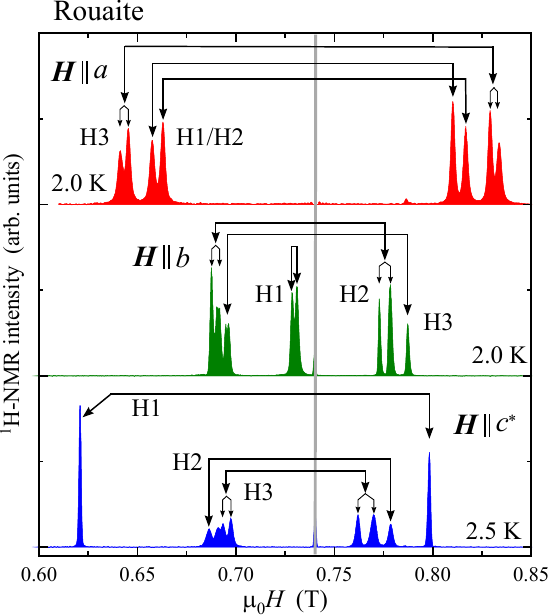}
\caption{
  Field-sweep $^1$H-NMR spectra at the lowest temperatures. 
  In the magnetically ordered state, each $^1$H-NMR peak splits into two lines due to internal field components parallel and anti-parallel to the external field direction. 
  In \Rou\ with three H sites, three pairs of peaks are observed. 
  \textcolor{\rev}{Large black arrows identify the pairs of split peaks. 
  A small extra splitting of peaks indicated by smaller arrows was observed due to a minor misalignment of the field direction from the crystal axes.} 
}
\label{SPABC}
\end{center}
\end{figure}

The NMR intensity is significantly reduced below the main AFM transition temperature $T_\text{N1}= 7.2$\,K.
This is attributable to the incommensurate magnetic structure in this intermediate temperature range \cite{Chakkingal-PRB110}. 
\textcolor{\rev}{The expanded spectrum at 5 K in $H\parallel a$ is shown together for a better visibility of small peaks observed in the incommensurate state. 
The peaks in the incommensurate state are assigned to the small part of the sample close to an impurity or surface, at which the magnetic structure is locked to the periodicity commensurate to the lattice. }
The temperature range corresponding to the incommensurate state at the measurement field is indicated by vertical arrows in Fig.~\ref{AFMSP}. 
We find that the lower transition temperature \textcolor{\rev}{$T_\text{N2}$} depends on the field orientation, which is consistent with the magnetic phase diagram constructed from bulk physical-properties measurements \cite{Chakkingal-PRB110}. 
In particular, the transition temperature increases when the field is along the $\bm{b}$ direction, while it decreases for other field orientations. 
When the NMR intensity is restored in the commensurate phase below the lower transition temperature $T_\text{N2}\approx 4$\,K, many sharp peaks with sizable peak separations are observed.  
The peak separation shows only a slight temperature dependence below 4\,K. 
In a canonical AFM state, the ordered moment appears at the transition temperature and grows upon further cooling, generating a mean-field--type temperature dependence of the NMR peak separation. 
In contrast, the NMR peak separation in rouaite is already large at the lower transition temperature, suggesting that the ordered moment appears at the main transition temperature $T_\text{N1}=7.2$\,K. 
This result indicates that the lower-temperature transition is a magnetic structure transition, and is consistent with the small anomaly in heat capacity and the neutron-diffraction study \cite{Chakkingal-PRB110}. 

For the low-temperature commensurate AFM state, the neutron-diffraction study observed the magnetic propagation vector $\bm{Q} = (0.5,0,0)$ \cite{Chakkingal-PRB110}. 
The doubling of the unit-cell along the $\bm{a}$ direction results in 12 H sites in a magnetic unit cell, and thus up to 12 peaks would be observed in the NMR spectrum. 
Two H sites are generated by the 2-fold screw operation in the structural unit cell.
The NMR peaks from these sites will overlap when the external field is along high-symmetry directions. 
In Fig.~\ref{SPABC}, we identify \textcolor{\rev}{ more than} 6 peaks in each spectrum at 2 K for $\bm{H}\!\parallel\bm{a}$ and $\bm{H}\!\parallel\bm{b}$ and 2.5 K for $\bm{H}\!\parallel\bm{c}^{\ast}$, and assign them to pairs as indicated by the black arrows from the symmetric temperature dependence of peak positions, allowing us to estimate the internal fields at the three H sites.
The small splitting of some of the peaks, \textcolor{\rev}{indicated by small arrows in Fig.~\ref{SPABC},} is attributed to a slight misalignment of the crystal relative to the external field and is neglected in the following discussion.

\section{Discussion}
In a magnetically ordered state with an antiferromagnetic component, the internal-field components parallel and anti-parallel to the external field direction serve to split an NMR peak. 
The peak separation is thus twice the internal field strength. 
From the peak assignment in Fig.~\ref{SPABC}, the internal fields at three H sites along the three mutually orthogonal directions \textcolor{\rev}{$a$, $b$, and $c^{\ast}$} are estimated to be 
\begin{align}
\bm{B}^{\rm H1}_{\rm int} &= (78, 1.5, 89)\; {\rm mT}, \notag \\
\bm{B}^{\rm H2}_{\rm int} &= (78, 44, 45)\; {\rm mT}, \notag \\
\bm{B}^{\rm H3}_{\rm int} &= (95, 46, 38)\; {\rm mT}. \notag 
\end{align} 
The internal field strengths cannot be uniquely assigned to the H1, H2 and H3 sites only based on the spectra shown in Fig.~\ref{SPABC};  simulations based on the magnetic structure model as discussed later are required for these peak assignments.

\begin{figure}[t]
\begin{center}
\includegraphics[width=0.95\columnwidth]{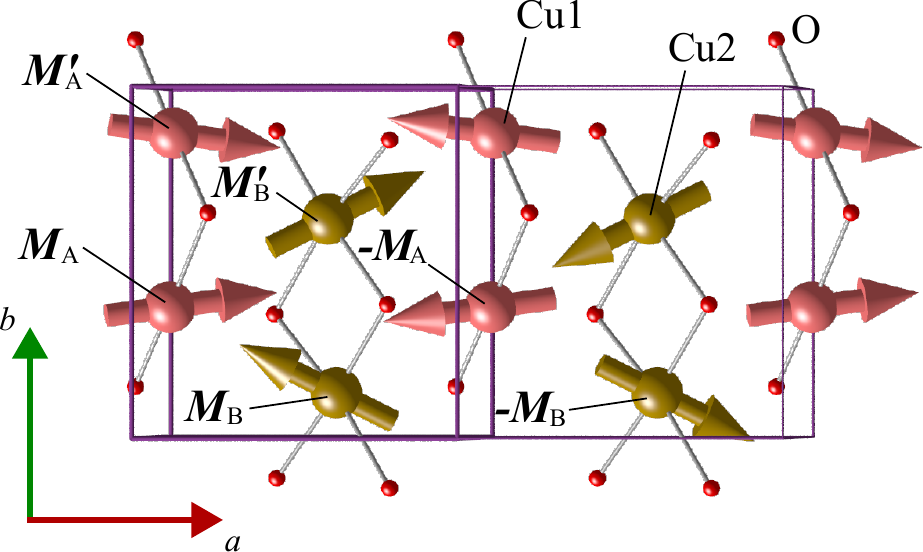}
\caption{
  Magnetic structure model for the internal-field analyses. 
  The magnetic unit cell is depicted by a thin purple line, while the crystallographic unit cell is indicated by a thicker purple line. 
  The magnetic moments on the Cu1 (Cu2) sites are labeled as $\bm{M}_{\rm A}$ ($\bm{M}_{\rm B}$). 
  For the FM Cu1 chain the magnetic moments are generated by the time-reversal operation after the $2_1$ screw operation along the chain. 
  For the AFM Cu2 chain, the time-reversal operation is omitted.
  2-D layers with the same magnetic structure are stacked along the $\bm{c}$ direction, leading to the magnetic propagation vector $\bm{Q}=(0.5,0,0)$. 
}
\label{magmodel}
\end{center}
\end{figure}

The internal fields at the H position were simulated based on the magnetic structure model suggested from the neutron diffraction study \cite{Chakkingal-PRB110}, where the Cu1 and Cu2 sites form ferromagnetic and antiferromagnetic chains along the $\bm{b}$ direction, respectively. 
As shown in Fig.~\ref{magmodel}, the magnetic unit cell is doubled along the $\bm{a}$ direction in the AFM state, so a translation along the $\bm{a}$ direction to the next structural unit cell inverts the direction of the magnetic moment. 
To model the magnetic structure, we assume that a magnetic moment at the Cu1 site is written as $\bm{M}^{\phantom{\prime}}_{\rm A} = (m^{\rm A}_a, m^{\rm A}_b, m^{\rm A}_{c^{\ast}})$ and at the Cu2 site as $\bm{M}^{\phantom{\prime}}_{\rm B} = (m^{\rm B}_a, m^{\rm B}_b, m^{\rm B}_{c^{\ast}})$. 
By applying the two-fold screw operation along the $\bm{b}$ direction, the magnetic moments adjacent to $\bm{M}^{\phantom{\prime}}_{\rm A}$ and $\bm{M}^{\phantom{\prime}}_{\rm B}$ are written as $\bm{M}'_{\rm A} =  (m^{\rm A}_a, -m^{\rm A}_b, m^{\rm A}_{c^{\ast}})$ and  $\bm{M}'_{\rm B} =  (-m^{\rm B}_a, m^{\rm B}_b, -m^{\rm B}_{c^{\ast}})$. 
Here, a time-reversal operation is applied in addition to the screw operation on the Cu1 chain to construct a ferromagnetic spin configuration. 
One example of this magnetic structure model is illustrated in Fig.~\ref{magmodel}.
The internal fields at the hydrogen sites are then simulated by calculating the hyperfine fields and dipole fields generated from these ordered moments.

The dipole coupling tensors for each Cu chain $\hat{D}_{\rm A}$ and $\hat{D}_{\rm B}$ are calculated using the magnetic structure model introduced above. 
The dipole fields $\bm{B}^{\rm dip}$ are obtained by the sum of both contributions
\begin{align}
\bm{B}^{\rm dip} = \hat{D}_{\rm A} \bm{M}_{\rm A} + \hat{D}_{\rm B} \bm{M}_{\rm B}.
\end{align} 
The two-fold screw operation inverts the sign of the $\bm{b}$ component for the H1 sites and the other two components for H2/H3 sites. 
To estimate the hyperfine fields, the Cu-site-dependent hyperfine coupling constants are required. 
However, experimentally obtained hyperfine coupling constants are the sum of all the contributions from the nearest neighbor Cu sites. 
We assumed equal contributions from the three neighboring  moments and divided the experimentally obtained coupling constants by three. 
\textcolor{\rev}{This approximation is coarse but reasonable because the hyperfine fields are compensated due to the alternating spin configuration, having anti-parallel spin components.
The effective hyperfine couplings are, thus, significantly reduced from the values listed in Table~\ref{tab:Ahf}, which are the maximum values for a FM spin configuration.}
In contrast, dipole fields are not canceled due to the offdiagonal components. 
With the diagonal hyperfine coupling tensor $\hat{A}$, the hyperfine fields at H1 and H2/H3 sites are calculated by the vector sum,  
\begin{align}
\bm{B}^{\rm hf}_1 &= \frac{\hat{A}}{3} \left( \bm{M}^{\phantom{\prime}}_{\rm A} + \bm{M}'_{\rm A} + \bm{M}^{\phantom{\prime}}_{\rm B} \right)  \notag \\ 
\bm{B}^{\rm hf}_{2,3} &= \frac{\hat{A}}{3} \left(\bm{M}^{\phantom{\prime}}_{\rm B} + \bm{M}'_{\rm B} + \bm{M}^{\phantom{\prime}}_{\rm A} \right).  \notag 
\end{align}
The hyperfine field is invariant against the two-fold screw operation but changes sign after translation along the $\bm{a}$ direction, reflecting the AFM structure. 

To explain the internal fields experimentally obtained at each H site and in three directions, $\bm{B}^{\rm dip} + \bm{B}^{\rm hf}$ are calculated for all possible combinations of $\bm{M}_{\rm A}$ and $\bm{M}_{\rm B}$. 
Each component of the magnetic moments is limited to $1.1\,\mu_{\rm B}$ in view of the realistic size of the ordered moments. 
The best sets of parameters are the following: 
\begin{align}
\bm{M}_{\rm A} &= \left( 0.64, 0.20, 0.72 \right)\mu_\text{B}, \notag \\
\bm{M}_{\rm B} &= \left( -0.60, 0.20, 0.62 \right)\mu_\text{B}. \notag
\end{align}
These results yield $|\bm{M}_{\rm A}| = 0.98\,\mu_{\rm B}$ and $|\bm{M}_{\rm B}| = 0.89\,\mu_{\rm B}$, suggesting that nearly full moments are ordered in the ground state, which is in good agreement with the neutron diffraction refinement. \cite{Chakkingal-PRB110} 
In the magnetic structure obtained, the internal fields at the H sites are estimated to be 
\begin{align}
\bm{B}^{\rm H1}_{\rm cal} &= (78, 2.2, 88)\; {\rm mT} \notag \\
\bm{B}^{\rm H2}_{\rm cal} &= (79, 44, 48)\; {\rm mT} \notag \\
\bm{B}^{\rm H3}_{\rm cal} &= (93, 45, 38)\; {\rm mT}. \notag 
\end{align} 
Our magnetic structure reproduces the experimental results almost perfectly, confirming our site assignment and the low-field magnetic structure of \Rou.

\begin{figure}[t]
\begin{center}
\includegraphics[width=0.9\columnwidth]{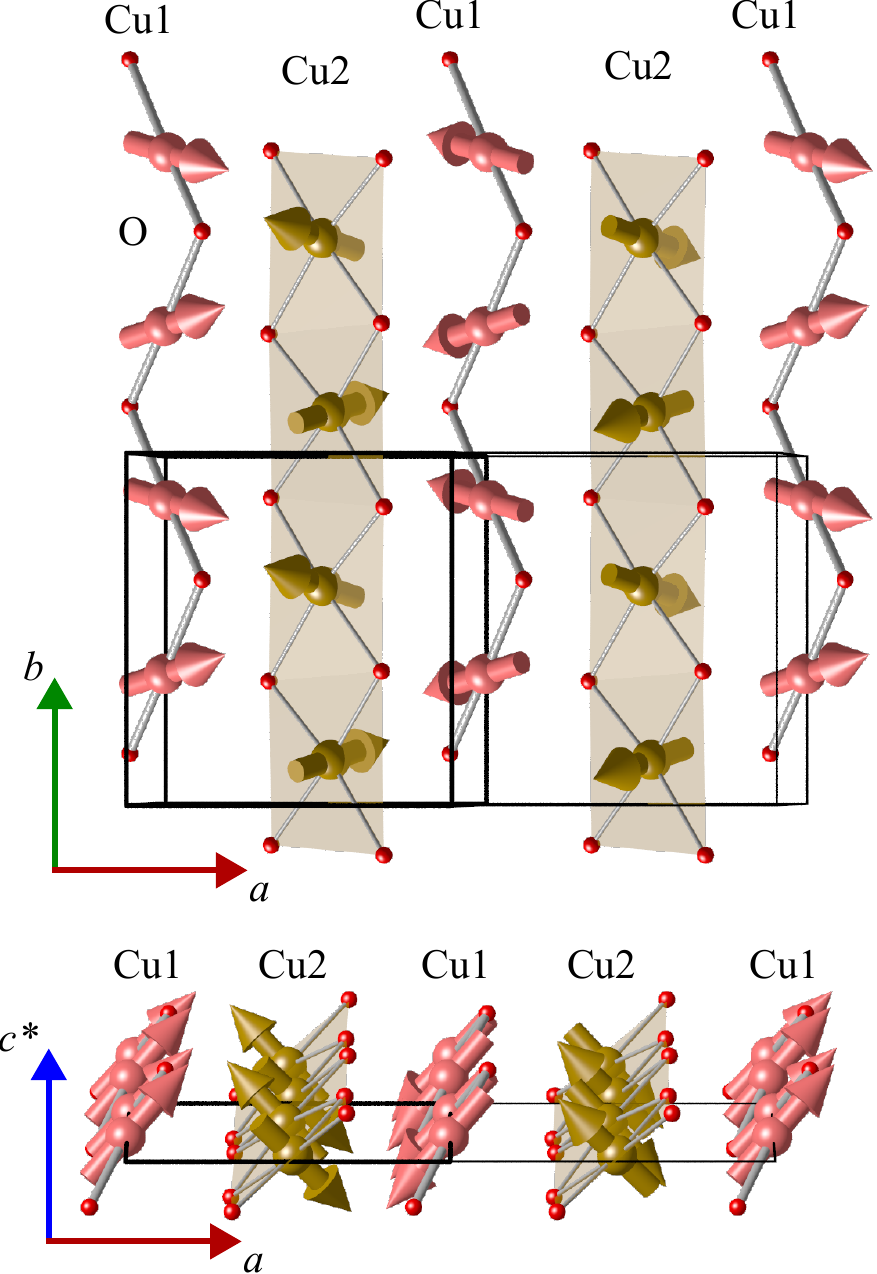}
\caption{
  The magnetic structure proposed by our analysis of the NMR spectra. 
  Both $\bm{M}_A$ and $\bm{M}_{\rm B}$ moments on Cu1 and Cu2 sites are canted along the chain direction. 
  The Cu2 moments are nearly perpendicular to the CuO$_4$ plaquette suggesting that the orientation of magnetic moments are pinned by the finite spin-orbit interactions. 
  The Cu1 moments are, then, perpendicular to the Cu2 moments. 
}
\label{LFstructure}
\end{center}
\end{figure}

The magnetic structure that best describes the present NMR measurements is illustrated in Fig.~\ref{LFstructure}. 
The ordered moments on both the FM and AFM chains are slightly canted along the $\bm{b}$ direction.
The finite $\bm{c}^{\ast}$ component leads to a zig-zag spin orientation along the $\bm{a}$ direction  [Fig.~\ref{LFstructure}(b)]. 
This magnetic structure is closely similar to that proposed previously based on neutron diffraction \cite{Chakkingal-PRB110}, but NMR is able to provide much more precise estimates of the canting angles, and this leads to two interesting insights.

First, it is noteworthy that the relative angles between $\bm{M}_{\rm A}$ and $\bm{M}_{\rm B}$ are close to 90$^{\circ}$.
Because the inter-chain exchange interactions are frustrated due to the mixture of FM and AFM interactions, the mutually orthogonal spin orientations likely serve to neutralize both the FM and AFM inter-chain interactions and to maximize the energy gain through the Dzyaloshinskii-Moriya (DM) interaction. 
Our results suggest that the DM vector has a finite component along the $\bm{b}$ direction. 

Second, as visualized in Fig.~\ref{LFstructure}(b), the magnetic moments at Cu2 sites lie perpendicular to the CuO$_4$ plane.
This suggests that the spin orientation at the Cu2 sites is influenced by spin-orbit (SO) interactions, while the alternating direction is determined by the exchange interactions. 
\textcolor{\rev}{The importance of the DM interactions suggested above is also supported by the finite SO interactions.}
It is surprising that the spin orientation would be locked perpendicular to the CuO$_4$ plaquettes by SO interactions, since relatively light Cu is not well known for strong SO coupling and $S=1/2$ quantum spins are also typically not SO coupled.  
In fact, the transition to an incommensurate phase at $\sim$\,4~K or 1--2\,T would indicate that the SO interaction is relatively weak.  
The weak SO interaction becomes important at low temperatures when the dominant but frustrated exchange interactions cannot select the most stable magnetic structure. 
In high magnetic fields, the FM chains are polarized first, overcoming the weaker, and frustrated, inter-chain interactions. 
The magnetic moments on the AFM chains will orient perpendicular to the external field, and thus to the orientation of the magnetic moments on the field-polarized FM chains. 

\begin{figure}[tb]
  \includegraphics[width=\columnwidth]{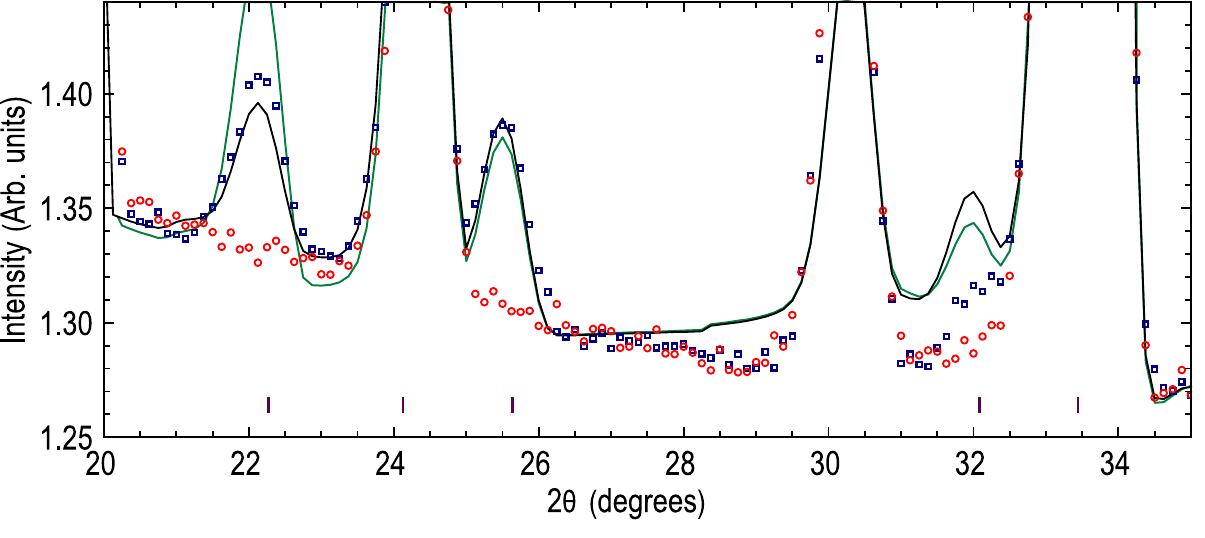}
  \caption{\label{neutron} Comparison of how well the refined magnetic structure (black line) and the magnetic structure determined from NMR (green line) describe the neutron diffraction data measured previously. \cite{Chakkingal-PRB110, ND_data} Red circle data were collected at 20\,K and blue square data at 3\,K; bars mark the positions of magnetic reflections.}
\end{figure}

Comparing Fig.~\ref{LFstructure} and the original magnetic structure refined by neutron diffraction \cite{Chakkingal-PRB110}, there are some significant differences in the canting angles, although the structures are qualitatively similar. 
The neutron diffraction refinement had significant uncertainties in the canting angles, so it is not immediately clear whether \textcolor{\rev}{the difference in canting angles exceeds experimental uncertainties}.  
We thus used the NMR-derived magnetic structure to simulate the neutron diffraction pattern, to indicate to what extent the canting angles are distinguishable by neutrons.
As shown in Fig.~\ref{neutron}, when the NMR magnetic structure is used to model the magnetic neutron diffraction pattern, the fit overestimates the intensity in the first magnetic peak, ($-\frac12$\,0\,1), which was previously underestimated, and leads to minor differences elsewhere.  
Aside from the first peak, it is not clear whether the new or previous magnetic orders better describe the diffracted intensity.  
The discrepancy in the ($-\frac12$\,0\,1) peak suggests that NMR sees stronger oscillation of the ordered moment along $\bm{a}$ than the neutrons do.  
One possible explanation for such a discrepancy is that the NMR is measured in fields that are roughly half that required to enter the incommensurate phase and one third to one quarter what would be required to flip one component of the spin along $\bm{a}$ or $\bm{c}^\ast$.
It is likely that the spins have rotated somewhat at these fields compared to the zero-field ground state.  
The extracted spin orientations are based on combining data from three orthogonal field directions, and that rotation due to the field will be different in each, but the rotation effect was not included in the NMR spectrum simulation. 
As such, it is unclear which set of canting angles is more plausible.  
Our main result, which seems robust, is that the NMR spectra are well described by a closely similar alternation of AFM and FM chains as in the neutron refinement, with similarly significant canting angles, and net moments along $\bm{b}$ on each AFM chain.  

\section{Conclusion}
The magnetic structure in the low-field AFM state of \Rou\ was revealed by measurement of $^1$H-NMR spectra for a single-crystalline sample. 
The directions of the ordered moments were determined by comparing the simulated internal fields at the target H sites and those estimated from the $^1$H-NMR spectra. 
We suggest that the orientations of the Cu2 moments are normal to the CuO$_4$ plaquettes fixed by the finite SO coupling. 
The competing inter-chain interaction is minimized by the orthogonal orientations between Cu1 and Cu2 chains. 
The external magnetic fields could modify the competing and weaker inter-chain interactions flipping selectively the magnetic moments on the FM Cu1 chains.
The present result proposes mechanisms of the low-field AFM structure, 
\textcolor{\rev}{which will be important to understand the 1/2 magnetization plateau observed in \Rou\, in higher fields \cite{Kikuchi-JPCS969,Yuan-PRB106}. 
Further microscopic studies in higher magnetic fields are required.}

\begin{acknowledgements}
This work was partially supported by JSPS KAKENHI (Grants Nos. 21H01035, 22H04458, 24H01599, and 25H00600). This work was also supported by the Deutsche Forschungsgemeinschaft (DFG, German Research Foundation) through individual Grants No.\ PE 3318/1-1 (Project No.\ 447786036) and PE 3318/2-1 (Project No.\ 452541981); project C03 of the Collaborative Research Center SFB 1143 (Project No.\ 247310070); and the W\"urzburg-Dresden Cluster of Excellence on Complexity and Topology in Quantum Materials\,---\,\textit{ct.qmat} (EXC 2147, Project No.\ 390858490). 
\end{acknowledgements}

\end{document}